\documentclass[11pt,a4paper]{amsart}
\usepackage{amssymb}

\newcommand{\NP}{Nucl. Phys. }
\newcommand{\PRL}{Phys. Rev. Lett. }
\newcommand{\Prev}{Phys. Rev. }
\newcommand{\PL}{Phys. Lett. }

\newcommand{\CMP}{Commun. Math. Phys. }

\newcommand{\MPL}{Mod. Phys. Lett. }

\def\RAMAN{\Psi}
\def\QPAR{q}
\def\QTIL{\bar q}
\def\QDILOG{e_b}
\def\FQDILOG{\phi}
\def\MOM{\mathsf P}
\def\POS{\mathsf X}

\def\COMPLEXS{\mathbb C}
\def\REALS{\mathbb R}
\def\INTEGERS{\mathbb Z}
\def\la{b}
\def\CLA{c_\la}
\def\IMUN{i}
\def\centc{c}

\def\mob{\triangleleft}
\def\f{\alpha}\def\g{\beta}\def\u{u}\def\p{u}
\def\l{L}\def\b{b}
\def\w{d}
\def\e{\epsilon}\def\x{f}
\title[strongly coupled quantum discrete  Liouville theory]
{Strongly coupled quantum discrete Liouville theory. I: Algebraic
  approach and duality} 
\author{L.D. Faddeev}
\address{Steklov Mathematical Institute at St. Petersburg,
Fontanka 27, St. Petersburg 191011, Russia\hfil\break
\indent Helsinki Institute of Physics, P.O. Box 9, FIN-00014
University of Helsinki, Finland}
\email{faddeev@pdmi.ras.ru, faddeev@pcu.helsinki.fi}
 \author{R.M. Kashaev}
\address{Steklov Mathematical Institute at St. Petersburg\hfil\break
\indent Helsinki Institute of Physics}
\email{kashaev@pdmi.ras.ru, kashaev@pcu.helsinki.fi}
 
\author{ A.Yu. Volkov}
\address{Steklov Mathematical Institute at St. Petersburg}
\curraddr{Vrije Universiteit Brussel
Pleinlaan 2 , B-1050 Brussel, Belgium}
\email{avolkov@vub.ac.be}

\date{May 2000}

\begin{document}
\begin{abstract}The quantum discrete Liouville model in the strongly 
  coupled regime, $1<c<25$, is formulated as a well defined quantum
  mechanical problem with unitary evolution operator. The theory is
  self-dual: there are two exponential fields related by Hermitean
  conjugation, satisfying two discrete quantum Liouville equations,
  and living in mutually commuting subalgebras of the quantum algebra
  of observables.
\end{abstract}
\maketitle
\section*{Introduction}
The Liouville equation\cite{liouville}
\begin{equation} \label{liouville}
\phi_{tt} -\phi_{xx}-4e^{-2\phi}=0  
        \end{equation}
        has plenty of important applications in Mathematics and
        Physics.  In particular, it describes the surfaces of constant
        negative curvature and plays the indispensable role in
        uniformization theory of Riemannian surfaces \cite{poinc} (for
        some recent approach see
        \cite{takhtajan,takhtajan1,takhtajan2}).
        
        In modern physics the Liouville equation defines one parameter
        family of models in conformal field theory (CFT), which
        usually is identified with 2-dimensional gravity
        \cite{jackiw}.  It plays even more important role in
        Polyakov's theory of noncritical Bosonic string in dimensions
        $d<26$ \cite{polyakov}. For these reasons the Liouville model,
        especially in its quantum version, attracted wide attention
        during the last 25 years \cite{curtho,dhoker,gernev1,gernev2}.
        
        The parameter $\centc$, labelling the quantum Liouville theory
        as a CFT model, is the central charge in a representation of
        the Virasoro algebra.  The quasiclassical (or weak coupling)
        region, corresponding to large positive $\centc$, is well
        understood. The domain $\centc<1$, containing the minimal
        models of CFT for some preferred discrete values of $\centc$,
        is also well described \cite{BPZ,fqs,tirkFad}. It is the
        region $1<\centc<25$, corresponding to strong coupling, about
        which there exists very limited knowledge until now.
        
        In this paper we begin to describe one more method to treat
        the Liouville model, which is applicable for studying the
        strong coupling region. The method is based on the apparatus
        of quantum integrable models (see e. g.  \cite{faddeev98} for
        the recent survey). Some parts of this machinery were already
        used in CFT and, in particular, for the Liouville model.
        However, we feel that we have added several new things to this
        development.
        
        First, we use the lattice regularization for the model, which
        exactly retains the integrability. Here we follow the previous
        papers \cite{fadtak,fadVol97}.  Second, we show that it is
        indispensable ( especially in the strong coupling region) to
        use simultaneously two mutually dual discrete models,
        corresponding to two exponents of the coupling constant
        $\tau^{\pm1}$, symmetrically entering the expression for the
        central charge
\[
\centc=1+6(\tau+\tau^{-1}+2).
\]
Positive $\tau$ correspond to $\centc>25$ or weak couplings, while
strong couplings lead to complex $\tau$ on unit circle,
\[
\tau=e^{\IMUN\theta},\quad \tau^{-1}=e^{-\IMUN\theta}=\bar\tau,
\]
and only unification of the models for $\tau$ and $1/\tau$ can restore
unitarity. Similar considerations on constructing ``modular doubles''
were used earlier in simpler examples of the Weil--Heisenberg algebra
\cite{faddeev95} and quantum group \cite{faddeev99}. This paper,
technically being more involved, ideologically belongs to the same
line of thought.  We must stress, that one can see the elements of the
dualization idea in papers \cite{babelon,gervais2,curtho,teschner}
devoted to the Liouville model itself, and in
\cite{BLZ1,BLZ2,BLZ3,BPZ} within the framework of conformal field
theory.

Our main result consists in constructing the unitary evolution
operator for the chiral shift serving the both dual models.

In the first part of the paper we remind various facts about the
Liouville equation to be used in what follows. Many of these facts are
known in one or another form; however we present them in the form most
suitable for our goal. Then follows description of the discretized
Liouville equation and its formal quantization in purely algebraic
manner.  The appropriate involution together with proper quantization
are introduced in section~\ref{sec:dual}. The self-dual structure is
indispensable for that.

This paper is partly supported by RFBR grant 99-01-00101, grant INTAS
99-01705, and Finnish Academy. A.V. wishes to acknowledge the
financial support extended by the DWTC office of the Belgian
government through the IUAP project P4/08. A.V. and L.D.F. are also
grateful to ESI, Vienna for hospitality during October, 1999.

\section{Recollecting the facts}
In this section we remind in appropriate form some facts about the
Liouville model and the technique of its discretization.
\subsection{Liouville formula and M\"obius-invariance}
Liouville formula makes solutions, all of them in fact, from pairs of
arbitrary functions of one variable ``moving'' in light-cone
directions
\begin{equation}\label{lf}
   -e^{-2\phi(x,t)}=\frac{\f'(x-t)\g'(x+t)}
   {(\f(x-t)-\g(x+t))^2}  \;  .      
\end{equation}
The right hand side of eqn~(\ref{lf}) is invariant under simultaneous
point-wise M\"obius transformations
\[ \f\mapsto\f\mob M=\frac{m_{21}+\f m_{11}}{m_{22}
  +\f m_{12}}\qquad\qquad\g\mapsto\g\mob M \] of the ``chiral
halves''~\(\f\) and~\(\g\).  Accordingly, if those are
``M\"obius-periodic''
\[ \f(x+2\pi)=\f(x)\mob T\qquad\qquad
\g(x+2\pi)=\g(x)\mob T , \] with the same monodromy matrix $T$ in both
chiralities of course, the solution comes out periodic in the spatial
direction
\[ \phi(x+2\pi,t)=\phi(x,t)    .                 \]
From now on, this will be the only boundary condition in use.

Because of this hidden M\"obius-symmetry, the chiral halves cannot be
uniquely restored from a given solution, but their Schwarz derivatives
can, for they are M\"obius-invariant as well. In particular, that of
\(\f\) (times minus one half)
\[ \textstyle\u=-\frac12\left(\frac{\f'''}{\f'}
  -\frac32(\frac{\f''}{\f'})^2\right) \] doubles as the (chiral
component of the) stress-energy tensor
\begin{equation}\label{se}\textstyle
   \u(x-t)=\frac14(\phi_x-\phi_t)^2+\frac12
   (\phi_{xx}-\phi_{tx})+e^{-2\phi} .     \end{equation}
We will now pay less attention the parallel, or
rather perpendicular, \(\g\)-chirality.
\subsection{Magri bracket}
The canonical Poisson bracket
\[ \{\varpi(x),\varphi(y)\}=\delta(x-y)          \]
on the laboratory Cauchy data
\[ \varphi=\phi|_{t=0}\qquad\qquad
\varpi=\phi_t|_{t=0} , \] together with canonical total momentum and
energy
\[ P=\int\varpi\varphi'dx\qquad\qquad
H={\tfrac12}\int(\varpi^2 +\varphi^{\prime2}+4e^{-2\varphi})dx \]
gives the equations of motion in the Hamiltonian formalism:
\begin{equation*}
   \begin{aligned} \varphi'&=\{P,\varphi\}       \\
     \varpi'&=\{P,\varpi\} \end{aligned} \qquad\qquad\begin{aligned}
     \dot\varphi&=\varpi=\{H,\varphi\}             \\
     \dot\varpi&=\varphi''+4e^{-2\varphi} =\{H,\varpi\}.
   \end{aligned}\end{equation*}
 These equations lead to a single free motion equation
\begin{equation}\label{cr}
   \dot\u=-\u'=\{\l_0,\u\}     ,     \end{equation}
with \(\l_0\) denoting the (slightly shifted)
zero'th Fourier coefficient
\[ \l_0=\int(\u+{\tfrac14})\:dx
={\tfrac12}(H-P+\pi) .  \] The bracket involved is the same canonical
bracket of course, but now we are prompted to express it, using
formula~(\ref{se}) at \(t=0\), in terms of either \(\u\) itself or its
Fourier coefficients \(\l_a=\int(\u+{\tfrac14})e^{-iax}dx\). The Magri
bracket \cite{magri}
\begin{equation}\label{mb}
   \{\u(x),\u(y)\}=(\u(x)+\u(y))\,\delta'(x-y)
   -{\tfrac12}\,\delta'''(x-y)       \end{equation}
and yet more celebrated (Poisson bracket
realization of) the Virasoro algebra
\[- i\{\l_a,\l_b\}=(a-b)\l_{a+b}
+\pi(a^3-a)\delta_{a,-b} \] emerge. They will attract our attention in
the next couple of (sub)sections where we shall try to treat the Magri
bracket and the ``chiral Hamiltonian''~\(\l_0\) as a stand-alone
dynamical system, not for the first time of course. We shall review
some earlier attempts to get it properly discretized and quantized. It
will give an idea how far we can go on bare common sense and formal
algebra, without actually touching the Liouville equation.

\subsection{Volterra model}
The first such attempt \cite{fadtak} produced a viable lattice
counterpart of the Magri bracket
\begin{equation*}\tag{\ref{mb}\('\)}\begin{aligned}
    \{\p_m,\p_n\}={\tfrac14}\:\p_m\p_n\big((4-\p_m
    &-\p_n)(\delta_{m+1,n}-\delta_{m-1,n})        \\
    &-\p_{\frac12(m+n)}(\delta_{m+2,n} -\delta_{m-2,n}) ,
  \end{aligned}\end{equation*} with \(\p_n\sim1-\Delta^2\u(n\Delta)\).
In the traditional difference-differential scheme of things it has to
do with Volterra's venerable preys and predators
\begin{equation*}\tag{\ref{cr}\('\)}
   \dot\p_n=\p_n(\p_{n-1}-\p_{n+1})=
   \left\{\sum\log\p_m,\p_n\right\},\end{equation*}
otherwise known as the lattice KdV equation. It
should be noted however that the latter only
emerges in a rather tricky continuous limit, while
the simplest limit leads to the free
equation~(\ref{cr}).
Which is a little unfortunate, for however
essential a part has the Volterra system been
playing in the soliton theory, there must be
something wrong about a nonlinear equation
emulating a linear one. Apparently, discrete space
and continuous time do not get along, still,
the KdV connection may teach us a thing or two.
\subsection{Bi-hamiltonity}
Historically, the Magri bracket brought about the notion of ``raising
and lowering'' which soon became paramount for the whole Hamiltonian
theory of soliton equation. In our case, one lifts~\(\l_0\) to
\[ \l_0^\uparrow=\int{\tfrac14}\:\u^2dx          \]
to turn the free motion~(\ref{cr}) into the KdV equation
\[ \dot\u={\tfrac12}\u'''-3\u\u'
=\{\l_0^\uparrow,\u\} , \] then drops the Magri bracket to that of
Zakharov-Faddeev's
\[ \{\u(x),\u(y)\}_\downarrow=2\delta'(x-y)      \]
to restore the free motion in a different guise
\[ \dot\u=-\u'=\{\l_0^\uparrow,\u\}_\downarrow . \]
Although not quite relevant to the Liouville equation, this makeshift
may be useful anyway, for the corresponding lower guise of the
Volterra system
\[ \dot\p_n=\p_n(\p_{n-1}-\p_{n+1})
=\left\{\sum\p_m,\p_n\right\}_\downarrow \] features a more
quantization-friendly bracket
\[ \{\p_m,\p_n\}_\downarrow=\p_m\p_n
(\delta_{m+1,n}-\delta_{m-1,n}).  \] We happen to already know
\cite{fadVol93} how to turn this one into a noncommutative algebra and
to arrange there a ``free motion in discrete time''. So, in the rest
of this section we shall be looking at that motion, then we shall try
to guess how it could be lifted back to the Magri--Virasoro case.
\subsection{Quantization}\label{subsec:goquant}
Going quantum, bracket \(_\downarrow\) ought to become Weyl-style
exchange relations
\[ \p_m\p_n=q^{2(\delta_{m+1,n}
  -\delta_{m-1,n})}\p_n\p_m\;.  \] We do not want to discuss before
time the nature of quantization constant~\(q\) or the exact degree of
formality implied in what follows, but we want to be very clear about
boundary conditions: the~\(\p\)'s are strictly periodic and so is the
Kronecker symbol of course
\[ \p_{n+N}=\p_n\qquad\qquad\delta_{m,n}
=\begin{cases}1&\text{if }m\equiv n\pmod{N}\\
  0&\text{otherwise,}\end{cases} \] on top of that, for the reason to
surface shortly, we want period~\(N\) to be even and an additional
condition
\begin{equation}
  \label{eq:c1c2}
\p_1\p_3\ldots\p_{N-1}=\p_2\p_4\ldots\p_N 
\end{equation}
to be met, it makes sense because the elements in both sides are
central. With all this in place, the following relation
\[ \hat\p_n=\p_{n-1}=Q^{-1}\p_nQ,                \]
with
\[ Q=\b_1\b_2\ldots\b_{N-1}\qquad\qquad
\b_n=\sum_{a=-\infty}^\infty q^{a^2}\p_n^a\;, \] proves to hold for
all~\(n\in\mathbb Z\). So, with automorphism
\(\hat{}:\p_n\mapsto\p_{n-1}\) interpreted as a jump in time, the
above relation replaces the (lower) Volterra system with a free and
fully discrete Heisenberg quantum motion, employing the element \(Q\)
as an ``evolution operator''.
\subsection{Proof}\label{subsecProof}
We just honestly start from \(n=1\):
\begin{multline*} Q^{-1}\p_1^{}Q=Q^{-1}\p_1^{}
  \b_1^{}\b_2^{}\b_3^{}\ldots\b_{N-1}^{}=Q^{-1}
  \b_1^{}\sum_{a=-\infty}^\infty q^{a^2+2a}\p_2^a
  \p_1^{}\b_3^{}\ldots\b_{N-1}^{}                \\
  =Q^{-1}\b_1^{}\b_2^{}q^{-1}\p_2^{-1}
  \p_1^{}\b_3^{}\ldots\b_{N-1}^{}=Q^{-1}\b_1^{}
  \b_2^{}\b_3^{}\p_2^{-1}\p_1^{}\p_3^{}\b_4^{}
  \ldots\b_{N-1}^{}=\cdots \end{multline*} --- every step adds another
factor, in the end parity of~\(N\) and that center-reducing
condition~(\ref{eq:c1c2}) make all the difference~---
\[ \cdots=Q^{-1}Q(\p_2^{}\p_4^{}\ldots
\p_{N-2}^{})^{-1}\p_1^{}\p_3^{} \ldots\p_{N-1}^{}=\p_{N}^{}.  \] We
may seem to have done only one very particular case, but, luckily, the
rest is little more than tautology. It immediately follows that
\(Q^{-1}\b_1Q=\b_N\), which can be written as
\[ Q=\b_1^{-1}Q\b_N^{}
=\b_2^{}\b_3^{}\ldots\b_N^{}, \] which in turn just means that
\(Q=\hat Q\) and therefore anything good for \(n=1\) at once becomes
just as good for all \(n\). That is it.

In the process, a seeming contradiction between the ``open-ended''
appearance of the evolution operator and the cyclic nature of its
action has resolved itself. Since
\[ Q=\b_1\b_2\ldots\b_{N-1}=\hat Q=\b_2\b_3\ldots
\b_N=\hat{\hat{Q}}=\cdots =\b_N\b_{N+1}\ldots\b_{N-2}\;, \] whichever
of these forms one chooses to express~\(Q\) in, it always remains an
ordered product but each time starts from another point, in this sense
it does not depend on the starting point, or rather does not have one.
The next subsection offers an explanation of this miracle.
\subsection{It has to do with braids}
Let us compile a list of what we know about the \(\b\)'s. They are
periodic, \(\b_{n+N}=\b_n\), they are this ``local''
\[ b_mb_n=b_nb_m\qquad
\text{if }|m-n|\not\equiv1\pmod{N} , \] and they satisfy ``global''
relations which closed the previous subsection
\[ \b_1\b_2\ldots\b_{N-1}=\b_2\b_3\ldots\b_N
=\cdots=\b_N\b_{N+1}\ldots\b_{N-2}\;.  \] Let us now consider these as
defining relations and identify the emerging group. It is \(B_N\), the
group of braids of \(N\) strings in 3 dimensions, alternatively
defined by an exhaustive list of (Artin's) relations
\[ \b_n\b_{n-1}\b_n=\b_{n-1}\b_n\b_{n-1}\qquad
\qquad\b_n\b_m=\b_m\b_n\mbox{ if }|n-m|\neq1 \] imposed on \(N-1\)
generators \(\b_1,\b_2,\ldots,\b_{N-1}\). We leave it as an
(instructive) exercise to check that our list is equivalent to
Artin's, provided the leftmost of our ``global'' relations is read as
\[ \b_N=(\b_2\ldots\b_{N-1})^{-1}
\b_1\b_2\ldots\b_{N-1} \] and understood as the definition of
\(\b_N\).  Just in case, let us warn against mistaking \(\b_N\) for
the \(N\)'th generator of \(B_{N+1}\), ours is just an element of
\(B_N\) explicitly defined above. Those familiar with the braid group
must have already recognized in it the first string crossing the very
last one behind all the others, let us picture it for \(N=4\):
\[ \setlength{\unitlength}{1mm}
\raisebox{4mm}{\(\b_1=\;\;\)}
\begin{picture}(12,11)\linethickness{0.3mm}
\qbezier(0,10)(-1,8)(1,6)\qbezier(3,4)(5,2)(4,0)
\qbezier(4,10)(4,6)(2,5)\qbezier(2,5)(0,4)(0,0)
\qbezier(8,10)(8,5)(8,0)\qbezier(12,10)(12,5)(12,0)
\end{picture}\qquad
\raisebox{4mm}{\(\b_2=\;\;\)} \setlength{\unitlength}{1mm}
\begin{picture}(12,11)\linethickness{0.3mm}
\qbezier(4,10)(3,8)(5,6)\qbezier(7,4)(9,2)(8,0)
\qbezier(8,10)(8,6)(6,5)\qbezier(6,5)(4,4)(4,0)
\qbezier(0,10)(0,5)(0,0)\qbezier(12,10)(12,5)(12,0)
\end{picture}\qquad
\raisebox{4mm}{\(\b_3=\;\;\)}
\begin{picture}(12,11)\linethickness{0.3mm}
\qbezier(8,10)(7,8)(9,6)\qbezier(11,4)(13,2)(12,0)
\qbezier(12,10)(12,6)(10,5)\qbezier(10,5)(8,4)(8,0)
\qbezier(0,10)(0,5)(0,0)\qbezier(4,10)(4,5)(4,0)
\end{picture}\qquad
\raisebox{4mm}{\(\b_4=\;\;\)} \setlength{\unitlength}{1mm}
\begin{picture}(12,11)\linethickness{0.3mm}
\qbezier(0,10)(-2,6)(1,5)
\qbezier[2](4,4)(7,3)(10,2)
\qbezier(10,2)(13,1)(12,0)
\qbezier(12,10)(13,8.5)(10,7.5)
\qbezier[2](10,7.5)(7,6.5)(4,5.5)
\qbezier(4,5.5)(1,4.5)(0,0)
\qbezier(4,10)(7,5)(4,0)\qbezier(8,10)(9,5)(8,0)
\end{picture}\raisebox{4mm}{ .}                  \]
A picture of the evolution operator
\[ \raisebox{4mm}{\(Q=\b_1\b_2\b_3=\b_2\b_3\b_4
  =\b_3\b_4\b_1=\b_4\b_1\b_2=\;\;\)} \setlength{\unitlength}{1mm}
\begin{picture}(12,11)\linethickness{0.3mm}
\qbezier(0,10)(0,8)(1,7.5)
\qbezier(11,2.5)(12,2)(12,0)
\qbezier(4,10)(4,9)(2,6)\qbezier(2,6)(0,3)(0,0)
\qbezier(8,10)(8,8)(6,5)\qbezier(6,5)(4,2)(4,0)
\qbezier(12,10)(12,7)(10,4)\qbezier(10,4)(8,1)(8,0)
\qbezier[3](1,7.5)(6,5)(11,2.5)
\end{picture}\raisebox{4mm}{ ,}                  \]
now explains better than words how it manages to be ordered and cyclic
at the same time.
\subsection{Lattice Virasoro algebra}\label{sec:lva}
Now we look for a generalization of this construction by natural
interpretation of the ``raising''.  So, we want the lattice Virasoro
algebra actually to be a group.  If that is to be, the ``global''
relations ought to remain the same as in the braid group
\[ \w_1\w_2\ldots\w_{N-1}=\w_2\w_3\ldots\w_N
=\cdots=\w_N\w_{N+1}\ldots\w_{2N-2}\;, \] or equivalently
\begin{equation*}\tag{\ref{cr}\(''\)}
   \hat\w_n=\w_{n-1}=Q^{-1}\w_nQ\qquad\qquad
   Q=\w_1\w_2\ldots\w_{N-1}\;    ,  \end{equation*}
but this time each generator should interfere not
only with the nearest neighbours but also with the
second-nearest ones
\[ \w_m\w_n=\w_n\w_m\qquad\text{if }|m-n|
\not\equiv1\text{ and }2\pmod{N} , \] just like it was in
bracket~(\ref{mb}\('\)) of course. This by itself implies
\[ \w_{n}\w_{n-2}\w_{n-1}\w_{n}\w_{n+1}
=\w_{n-2}\w_{n-1}\w_{n}\w_{n+1}\w_{n-1} \] instead of Artin's
relations, but we feel that these are not tight enough, so we
voluntarily split each of them in two
\[ \w_{n}\w_{n-2}\w_{n-1}\w_{n}\w_{n+1}=\w_{n-2}
\w_{n}\w_{n-1}\w_{n+1}=\w_{n-2}\w_{n-1}\w_{n} \w_{n+1}\w_{n-1} , \] to
end up with weird relations from \cite{volkov97c}
\begin{equation*}\tag{\ref{mb}\(''\)}
   \begin{aligned}\w_{n+1}\w_{n-1}\w_{n}\w_{n+1}
     &=\w_{n-1}\w_{n+1}\w_{n}                      \\
     \w_{n}\w_{n-1}\w_{n+1}&=\w_{n-1}\w_{n}\w_{n+1} \w_{n-1} .
   \end{aligned}\end{equation*} 
 This coup de force will find its justification in the treatment of
 the Liouville model below.
\section{Difference-difference Liouville equation}
\subsection{Liouville formula} For aesthetical
reason alone, its difference approximation can only be
\[ \begin{picture}(260,65)(-10,0)\thicklines
\put(-10,5){\vector(1,0){85}}\put(80,2){\(x\)}
\put(0,0){\vector(0,1){40}}\put(-10,37){\(t\)}
\put(10,20){\line(1,1){40}}\put(53,57){\(\f\)}
\put(10,50){\line(1,-1){40}}\put(0,47){\(\g\)}
\put(20,10){\line(1,1){40}}\put(63,47){\(\f'\)}
\put(20,60){\line(1,-1){40}}\put(10,57){\(\g'\)}
\put(31,32){\(\phi\)}\put(120,32){\(-\Delta^2
e^{-2\phi}\sim\dfrac{(\f'-\f)
(\g'-\g)}{(\f'-\g')(\f-\g)}\;,\)} \end{picture}  \]
with primes now denoting finite shifts of arguments
by \(\Delta\). Appearance aside, we have already
mentioned that the Liouville formula manifests
invariance of \(\phi\) under simultaneous
point-wise M\"obius transform of chiral
halves~\(\f\) and~\(\g\). The so called cross-ratio
employed in our difference scheme has just
the same property.

To make of this a lattice formula we draw a square \(j,k\) lattice and
put
\[ \chi_{jk}=-\frac{\left(\f_{\frac12(j-k+1)}
    -\f_{\frac12(j-k-1)}\right) \left(\g_{\frac12(j+k+1)}
    -\g_{\frac12(j+k-1)}\right)} {\left(\f_{\frac12(j-k+1)}
    -\g_{\frac12(j+k+1)}\right) \left(\f_{\frac12(j-k-1)}
    -\g_{\frac12(j+k-1)}\right)} \] in the vertices with
\(j+k\equiv1\pmod2\), like this:
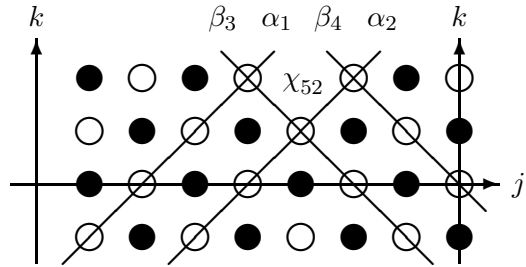
\[ \begin{picture}(195,100)(30,-30)\thicklines
\put(30,0){\vector(1,0){185}}
\put(220,-3){\(j\)}
\put(40,-30){\vector(0,1){85}}
\put(37,60){\(k\)}
\put(200,-30){\vector(0,1){85}}
\put(197,60){\(k\)}
\put(50,-30){\line(1,1){80}}
\put(125,60){\(\f_1\)}
\put(90,-30){\line(1,1){80}}
\put(165,60){\(\f_2\)}
\put(110,50){\line(1,-1){80}}
\put(105,60){\(\g_3\)}
\put(150,50){\line(1,-1){60}}
\put(145,60){\(\g_4\)}
\put(60,-20){\circle{10}}\put(60,0){\circle*{10}}
\put(60,20){\circle{10}}\put(60,40){\circle*{10}}
\put(80,-20){\circle*{10}}\put(80,0){\circle{10}}
\put(80,20){\circle*{10}}\put(80,40){\circle{10}}
\put(100,-20){\circle{10}}\put(100,0){\circle*{10}}
\put(100,20){\circle{10}}\put(100,40){\circle*{10}}
\put(120,-20){\circle*{10}}\put(120,0){\circle{10}}
\put(120,20){\circle*{10}}\put(120,40){\circle{10}}
\put(140,-20){\circle{10}}\put(140,0){\circle*{10}}
\put(140,20){\circle{10}}
\put(133,37){\(\chi_{52}^{}\)}
\put(160,-20){\circle*{10}}\put(160,0){\circle{10}}
\put(160,20){\circle*{10}}\put(160,40){\circle{10}}
\put(180,-20){\circle{10}}\put(180,0){\circle*{10}}
\put(180,20){\circle{10}}\put(180,40){\circle*{10}}
\put(200,-20){\circle*{10}}\put(200,0){\circle{10}}
\put(200,20){\circle*{10}}\put(200,40){\circle{10}}
   \end{picture}                                 \]
   So, the \(\chi\)'s occupy sites marked by bullets, the empty
   circlets will this time remain empty. The second \(k\)-axis at
   \(j=8\) reminds that the lattice covers a cylinder rather than a
   plane, \(\chi_{j+2N,k}=\chi_{jk}\) , with \(N\) not necessarily
   equal four of course.
\subsection{Discrete Liouville equation}
It not only ideologically justifies the chosen discretization but also
helps make the following action more entertaining. In continuum, the
Liouville equation may be considered as a compatibility condition for
the Liouville formula overloaded with M\"obius-invariance. It ought to
be just the same on the lattice, so let us find out what ``equation of
motion'' the \(\chi\)'s might solve. We pick four of them next to each
other to figure out that they are four cross-ratios made from the
total of six~\(\f\)'s and~\(\g\)'s. That is one too many: 4~\(\chi\)'s
+ 3~symmetry parameters -- 3~\(\f\)'s -- 3~\(\g\)'s = 1. The missing
one turns out to be
\begin{equation} \label{lle}
   \chi_{j,k+1}\chi_{j,k-1}=\frac{\chi_{j-1,k}
   \chi_{j+1,k}}{(1+\chi_{j-1,k})
   (1+\chi_{j+1,k})}\qquad j+k\equiv0\!\!
   \pmod2\!\!\!\!\!\!\!\!\!          \end{equation}
which so becomes our favourite lattice Liouville
equation --- replacing Hirota's original
\[ h_{j,k+1}h_{j,k-1}=\frac{h_{j-1,k}h_{j+1,k}}
{1+h_{j-1,k}h_{j+1,k}} .  \] The two are connected by a simple change
of variables
\[ \chi_{jk}=h_{j-1,k}h_{j+1,k}    ,             \]
which quarters the \(h\)'s in the empty sites of our lattice of
course. However, that change would also split cross-ratios, which we
want to avoid, at least in this paper.
 
The stress-energy connection~(\ref{se}) also lends itself to a
cross-ratio treatment. The Schwarz derivative inevitably becomes
another cross-ratio \cite{fadtak1}
\[ \p_n=4\frac{(\f_{n+2}-\f_{n+1})
  (\f_{n}-\f_{n-1})}{(\f_{n+2} -\f_{n-1})(\f_{n+1}-\f_{n-1})} , \]
then another counting argument leads to
\begin{equation} \label{lse}
   \frac4{\p_{\frac12(j-k)}}=\left(1+\chi_{j,k-1}
   +\dfrac{\chi_{j,k-1}}{\chi_{j-1,k}}\right)
   \left(1+\chi_{j+1,k}+\dfrac{\chi_{j+1,k}}
   {\chi_{j+2,k-1}}\right) \;.       \end{equation}
\subsection{Poisson bracket}
Two rows of \(\chi\)'s make perfect Cauchy data
\[ \chi_{2n}=\chi_{2n,-1}\qquad\qquad
\chi_{2n+1}=\chi_{2n+1,0}\;, \] just in case one may directly check
that the lattice canonical bracket of \cite{fadVol97}
\[ \{\chi_{2n\pm1},\chi_{2n}\}=\chi_{2n\pm1}
\chi_{2n}\qquad\{\chi_j,\chi_i\}=0 \mbox{ if }|j-i|\not\equiv1\pmod{N}
\] both a) reproduces itself as the \(\chi\)'s evolve according to the
lattice Liouville equation (\ref{lle}) and b) translates, by means of
formula~(\ref{lse}) at \(k=0\), into the lattice Magri
bracket~(\ref{mb}\('\)).
\section{Formal quantization}\label{sec:fquant}
\subsection{Algebra of observables}
In a virtual replay of subsection~\ref{subsec:goquant}, we introduce a
formal quantization constant \(q\), replace the above lattice
canonical bracket by Weyl-style exchange relations
\[ \chi_{2n\pm1}\chi_{2n}=q^2\chi_{2n}
\chi_{2n\pm1}\qquad\chi_j\chi_i=\chi_i\chi_j \mbox{ if
  }|j-i|\not\equiv1\pmod{2N} , \] opt for even \(N\) and impose an
additional condition on central elements
\begin{equation}\label{cc}
   \chi_1^{}\chi_3^{-1}\chi_5^{}\chi_7^{-1}\ldots
   \chi_{2N-3}^{}\chi_{2N-1}^{-1}=\chi_2^{}
   \chi_4^{-1}\chi_6^{}\chi_8^{-1}\ldots
   \chi_{2N-2}^{}\chi_{2N}^{-1}\;.   \end{equation}
\subsection{Quantum lattice Liouville equation}
We can only afford a Heisenberg evolution of observables, so let us
again consider \(\chi_j\) as initial data (\(\chi_{2n,-1}=\chi_{2n}\),
\(\chi_{2n+1,0}=\chi_{2n+1}\)) and determine elements
\(\chi_{j,k+1}\), \(j+k\equiv0\pmod2\) step by step using a slightly
``quantized'' lattice Liouville equation
\begin{equation} \label{qeq}
   \chi_{j,k+1}\chi_{j,k-1}=\frac{q^2\chi_{j-1,k}
   \chi_{j+1,k}}{(1+q\chi_{j-1,k})
   (1+q\chi_{j+1,k})}    .           \end{equation}
We deliberately present the r.h.s. as a ratio
to stress that all the factors there commute with
each other, on the contrary, those in the l.h.s
commute neither with each other nor with the
r.h.s..
\subsection{Shift and evolution operators}\label{subsec:shevol}
If we guessed the equation right, there must exist the ``evolution
operator'', that is an element \(K\) such that
\[ \chi_{j,k+1}=K^{-1}\chi_{j,k-1}K.             \]
Needless to say, these relations will hold everywhere if and only if
they do so for \(k\) equal \(0\) and \(-1\) where they become
\begin{align*} \chi_{2n-1}K\chi_{2n-1}
  K^{-1}&=\frac{q^2\chi_{2n-2}\chi_{2n}}
  {(1+q\chi_{2n-2})(1+q\chi_{2n})}              \\
  K^{-1}\chi_{2n}K\chi_{2n} &=\frac{q^2\chi_{2n-1}\chi_{2n+1}}
  {(1+q\chi_{2n-1})(1+q\chi_{2n+1})}. \end{align*} Following
\cite{fadVol93,fadVol97} we shall produce that element, with
explicitly singled out d'Alembert part \(K_\infty^{}\) good for
\[ \chi_{2n-1}K_\infty^{}\chi_{2n-1}
K_\infty^{-1}=q^2\chi_{2n-2}\chi_{2n}\qquad
K_\infty^{-1}\chi_{2n}K_\infty^{}\chi_{2n} =q^2\chi_{2n-1}\chi_{2n+1}
, \] and complete with the ``shift operator''~\(J\) moving
the~\(\chi\)'s in the spatial direction
\[ \chi_{j+1,k}=J^{-1}\chi_{j-1,k}J  .           \]
Here follow explicit formulas for these shift and evolution operators
\[ K_\infty^{}=VU\qquad\qquad K=E_2K_\infty^{}E_1
\qquad\qquad J=VU^{-1} \] where
\begin{align*} E_1&=\prod_{n=1}^N
  \e(\chi_{2n-1})\qquad\qquad
  E_2=\prod_{n=1}^N\e(\chi_{2n})                \\
  U&=\theta(q\chi_1^{-1}\chi_2^{}) \theta(q\chi_3^{-1}\chi_4^{})\ldots
  \theta(q\chi_{2N-3}^{-1}\chi_{2N-2}^{})       \\
  V&=\theta(q\chi_{2N-1}^{}\chi_{2N-2}^{-1})
  \theta(q\chi_{2N-3}^{}\chi_{2N-4}^{-1})\ldots
  \theta(q\chi_3^{}\chi_2^{-1}) .  \end{align*} Of course, \(U\) and
\(V\) have everything to do with braids but we will not go into that.
The two special functions involved
\begin{equation*}\begin{aligned}
    \e(z)&=\prod_{p=0}^\infty(1+q^{2p+1}z)
    =(-qz;q^2)_\infty                             \\
    \theta(z)&=\e(z)\,\e(z^{-1})=\text{const}
    \cdot\sum_{p=-\infty}^\infty q^{p^2}z^p
                       \end{aligned}\end{equation*}
                     are quite special indeed but all we want to know
                     about them, for now at least, are the beautiful
                     functional equations
\[ \frac{\e(qz)}{\e(q^{-1}z)}=\frac1{1+z} \qquad\qquad  
\frac{\theta(qz)} {\theta(q^{-1}z)}=\frac1{z}\; \] fulfiled by the
former, and the ensuing equation on the latter which has already been
used, albeit implicitly, in subsection~2.4.  This time those
functional equations and condition (\ref{cc}) gradually translate into
relations
\begin{xxalignat}{2} E_1^{-1}\chi_{2n-1}^{}E_1
  =\chi_{2n-1}^{}&&&E_1^{-1}\chi_{2n}^{}E_1 =\chi_{2n}^{}
  (1+q\chi_{2n-1}^{})(1+q\chi_{2n+1}^{})        \\
  E_2^{-1}\chi_{2n}^{}E_2 =\chi_{2n}^{}&&&E_2^{-1}\chi_{2n-1}^{}E_2
  =((1+q\chi_{2n-2}^{})
  (1+q\chi_{2n}^{}))^{-1}\chi_{2n-1}^{}         \\
  U^{-1}\chi_{2n}^{}U=\chi_{2n-1}^{}&&&
  U^{-1}\chi_{2n+1}^{}U=q^2\chi_{2n-1}^{}
  \chi_{2n+1}^{}\chi_{2n}^{-1}                  \\
  V^{-1}\chi_{2n}^{}V=\chi_{2n+1}^{}&&&
  V^{-1}\chi_{2n-1}^{}V=q^2\chi_{2n-1}^{} \chi_{2n+1}^{}\chi_{2n}^{-1}
\end{xxalignat} which combined make~\(J\) and~\(K\) satisfy what they
have to. We omit the calculation but a remark is in order.
\subsection{Odd \(N\) or no condition~(\ref{cc})}
In these cases it all fails, and subsection~\ref{subsecProof} gives an
idea why. In fact, they are better served by those
cross-ratio-splitting variables \(h\) and Hirota's lattice Liouville
equation which we were so quick to discard. We are planning to return
to this issue elsewhere.
\subsection{Chiral evolution operator}
We define it as
\[ Q=UE_1   \;   ,                               \]
and it indeed moves the \(\chi\)'s in the right direction
\[ \chi_{j,k+1}=Q^{-1}\chi_{j+1,k}Q              \]
and equals~\(\sqrt{J^{-1}K}\), in the sense that \(Q^2=J^{-1}K\). So,
the total momentum~\(P\), Hamiltonian~\(H\) and the chiral
Hamiltonian~\(\frac12(H-P)\) of the original equation have finally
become shift-evolution-operators~\(J\),~\(K\) and~\(Q\) of the quantum
lattice equation, but we have yet to find out in what sense, if at
all, this~\(Q\) coincides with that hypothetical quantum lattice
group-like counterpart of~\(L_0\) which we called~\(Q\) in
section~\ref{sec:lva}.  The answer does not come easy but in the end
few carefully placed~\(q\)'s do it again, and so does the magic
function~\(\e\) introduced in subsection~\ref{subsec:shevol}.  It
turns out that
\begin{equation}
  \label{eq:qdd}
Q=\w_1\w_2\ldots\w_{N-1}      ,      
\end{equation}
where~\(Q\) is~\(UE_1\) and the~\(\w\)'s are
\[ \w_n=\e\big(q^{-1}(1+q\chi_{2n}^{}+\chi_{2n}^{}
\chi_{2n-1}^{-1})(1+q\chi_{2n+1}^{}
+\chi_{2n+2}^{-1}\chi_{2n+1}^{})-q^{-1}\big). \] As the notation
suggests, these also satisfy relations~(\ref{cr}\(''\))
and~(\ref{mb}\(''\)). It is instructive now to see, that argument of
function $\epsilon$ in this formula is a natural quantisation of
expression (\ref{lse}).

\subsection{Proof of eqn~(\ref{eq:qdd})}
First, we recall the two identities satisfied by $\epsilon$-function:
\begin{equation}
  \label{eq:idep}
\e(u)\e(v)=\e(u+v),\quad \e(v)\e(u)=\e(v+u+qvu),\quad uv=q^2vu .
\end{equation}
Next, we have the identity
\[
\e(\chi_{1}(1+q\chi_{2}^{-1}))Q=Q
\e(\chi_{2N}^{-1}(1+q\chi_{2N-1}^{-1})^{-1})
\]
which is a consequence of the relations in
subsection~\ref{subsec:shevol} satisfied by $U$ and $E_1$ operators.
It is also easily checked that for any $j$
\[
\theta(q\chi_{2j-1}^{-1}\chi_{2j})\e(\chi_{2j-1})=
\e(\chi_{2j-1}(1+q\chi_{2j}^{-1})) \e((1+q\chi_{2j-1}^{-1})\chi_{2j}).
\]
Now, using these relations, we have
\begin{multline*}
  Q=(\e(\chi_{1}(1+q\chi_{2}^{-1})))^{-1}UE_1\e(\chi_{2N}^{-1}(1+q\chi_{2N-1}^{-1})^{-1})\\
  =(\e(\chi_{1}(1+q\chi_{2}^{-1})))^{-1}\left(\prod_{1\le j<N}
    \theta(q\chi_{2j-1}^{-1}\chi_{2j})\e(\chi_{2j-1})\right)\\
  \times\e(\chi_{2N-1}) \e(\chi_{2N}^{-1}(1+q\chi_{2N-1}^{-1})^{-1})\\
  =(\e(\chi_{1}(1+q\chi_{2}^{-1})))^{-1}\left(\prod_{1\le j<N}
    \e(\chi_{2j-1}(1+q\chi_{2j}^{-1}))
    \e((1+q\chi_{2j-1}^{-1})\chi_{2j})\right)\\
  \times\e(q\chi_{2N-1}\chi_{2N}^{-1})\e(\chi_{2N-1}) \\
  =\prod_{1\le j<N} \e((1+q\chi_{2j-1}^{-1})\chi_{2j})
  \e(\chi_{2j+1}(1+q\chi_{2j+2}^{-1}))=\prod_{1\le j<N}\w_j,
\end{multline*}
thus obtaining eqn~(\ref{eq:qdd}).

The proof of relations (\ref{mb}\(''\)) is done in \cite{volkov97c} and
is also based on identities (\ref{eq:idep}).

\section{Dualization}\label{sec:dual}
\subsection{Involution} The formal quantization of
section~\ref{sec:fquant} can be put in the usual framework of quantum
mechanics if we assume that both the formal parameter $q$ and the
generators \(\chi\) are exponentials
\[ q=e^{\pi i\tau}\qquad\qquad
\chi_j=e^{-2\pi\sqrt\tau\varphi_j}, \] and the new generators
$\varphi_j$ have the commutation relations
\[ [\varphi_{2n\pm1},\varphi_{2n}]=-\tfrac{I}
{2\pi i}\qquad\qquad[\varphi_j,\varphi_i]=0 \mbox{ if
  }|j-i|\not\equiv1\pmod{N} \] independent of $\tau$, and so could be
taken as selfadjoint operators
\[ \varphi_j^\dagger=\varphi_j^{} .                  \]
The rest depends on~\(\tau\) of course.  If the established formula
\[ c=1+6(\tau+\tfrac1{\displaystyle\tau}+2)      \]
(relating the coupling constant of the continuous Liouville field
theory to the central charge of the corresponding representation of
the Virasoro algebra) has something to do with our lattice theory,
then three cases have to be considered: a)
\(c\leq1\leftrightarrow\tau<0\); b) \(c\geq25\leftrightarrow\tau>0\);
and c) \(1\leq c\leq25\leftrightarrow|\tau|=1\).  The first two,
although lead to pretty normal reality conditions
\(\chi_j^\dagger=\chi_j^{-1}\) and \(\chi_j^\dagger=\chi_j^{}\)
respectively, also put~\(q\) on the unit circle, which is the last
thing we want right now. That leaves c).
\subsection{Change of function $\e$ } So, \(|\tau|=1\), also,
since \(c\) does not distinguish between \(\tau\) and \(1/\tau\), let
for definiteness \(\Im\tau>0\).  Following~\cite{faddeev95}, consider
the function
\begin{equation}
  \label{eq:fofz}
\x(z)=e_{\sqrt\tau}(\IMUN z/\sqrt\tau
)=\frac{(-e^{2\pi i(z+\frac\tau2)};
  e^{2\pi i\tau})_\infty} {(-e^{\frac{2\pi i}\tau(z-\frac12)};
  e^{-\frac{2\pi i}\tau})_\infty} \; ,  
\end{equation}
which uses our former favourite~\(\e\) as the numerator but divides it
by itself but with suitably altered arguments. In Appendix we collect
some of the properties of this function. Here we remark in particular
that \(\x(z)\) satisfies the same functional equation
\begin{equation}\label{fe}
   \frac{\x(z+\frac\tau2)}{\x(z-\frac\tau2)}
   =\frac1{1+e^{2\pi i\tau z}}       \end{equation}
as~\(\e(e^{2\pi iz})\) did, but this time we
also have
\[ |\x(z)|=1                                     \]
on the line~\(z=\tau\bar{z}\), simply because the numerator and
denominator of~\(\x\) are complex conjugate of each other on that
line.  There is an easy profit to be had from that.
\subsection{Unitarity}
Replace~\(\e\) by~\(\x\) in every factor of every shift-evolution
operator, all those little factors and big operators will at once
become unitary but all the relations we have found them to satisfy
will remain intact, for they rely only on that one and only functional
equation. For instance, let us write out the so upgraded chiral
evolution operator:
\[ Q=\kappa^{2(N-1)}
e^{\pi i(\varphi_1-\varphi_2)^2} e^{\pi
  i(\varphi_3-\varphi_4)^2}\ldots e^{\pi
  i(\varphi_{2N-3}-\varphi_{2N-2})^2}
\prod_{n=1}^N\x(i\sqrt\tau\varphi_{2n-1}) , \] where it is already
taken into account that
\[ \x(z)\x(-z)=\kappa^2e^{-\frac{\pi i}\tau z^2},\]
with~\(\kappa=\x(0)\) of course.

Mission accomplished, our lattice Liouville model has finally turned
from an algebraic fantasy into a quite material unitary theory.
Moreover, we shall momentarily see that we get a new feature.
\subsection{Dual Liouville equation}
Let us permute the factors in the l.h.s of the lattice Liouville
equation~(\ref{qeq})
\begin{equation*}\begin{aligned}
    \chi_{j,k-1}^{}\chi_{j,k+1}^{}
    &=\chi_{j,k-1}^{}\left(\chi_{j,k+1}^{}
      \chi_{j,k-1}^{}\right)\chi_{j,k-1}^{-1}       \\
    &=\chi_{j,k-1}^{}\left(\frac{q^2\chi_{j-1,k}
        \chi_{j+1,k}}{(1+q\chi_{j-1,k})
        (1+q\chi_{j+1,k})}\right)\chi_{j,k-1}^{-1}    \\
    &=\frac{q^{-2}\chi_{j-1,k}\chi_{j+1,k}}
    {(1+q^{-1}\chi_{j-1,k})(1+q^{-1}\chi_{j+1,k})}
                       \end{aligned}\end{equation*}
                     and treat the both sides with Hermitean
                     conjugation, remembering that
                     \(Q^\dagger=Q^{-1}\) of course.  Since
\[\tilde q \equiv \overline{q^{-1}}=e^{\frac{\pi i}\tau}
=q^{(\frac1\tau)^2}\qquad\qquad \tilde\chi_j\equiv\chi_j^\dagger
=e^{-\frac{2\pi} {\sqrt\tau}\varphi_j}=\chi_j^{\frac1\tau} , \] the
resulting equation reads
\begin{equation*}\tag{\ref{qeq}\(^\ast\)}
   \tilde\chi_{j,k+1}
   \tilde\chi_{j,k-1}
   =\frac{\tilde q^2
   \tilde\chi_{j-1,k}
   \tilde\chi_{j+1,k}}
   {(1+\tilde q
   \tilde\chi_{j-1,k})
   (1+\tilde q
   \tilde\chi_{j+1,k})}   .  \end{equation*}
It is plain to see that it is the same equation
but with~\(q\) and~\(\chi\)'s
replaced by~\(\tilde q\)
and~\(\tilde\chi\)'s,
this is called duality. Indeed,
instead of conjugating things, we might instead
explore the fact that
those~\(\tilde\chi\)'s satisfy the same
relations
\[ \tilde\chi_{2n\pm1}\tilde\chi_{2n}
=\tilde q^2 \tilde\chi_{2n} \tilde\chi_{2n\pm1} \] as the
original~\(\chi\)'s do, except with~\(\tilde q\) instead of~\(q\).
Moreover, the two sets commute with each other
\[ \chi_j^{}\tilde\chi_i
=\tilde\chi_i\chi_j^{}\;.  \] The algebras they generate (call
them~\(\mathcal A_\tau\) and~\(\mathcal A_{\frac1\tau}\)) form two
factors in the algebra~\(\mathcal B\) generated by the~\(\varphi\)'s,
and leave no free space, in the sense that \(\mathcal B=\mathcal
A_\tau \otimes\mathcal A_{\frac1\tau}\), sort of. Either way, such a
bisection is well served by the function~\(\x\), which fittingly
satisfies the dual functional equation
\begin{equation*}\tag{\ref{fe}\(^\ast\)}
   \frac{\x(z+\frac12)}{\x(z-\frac12)}=\frac1
   {1+e^{\frac{2\pi i}\tau z}}      \end{equation*}
to complement the original one~(\ref{fe}).
Equation~(\ref{qeq}\(^\ast\)) can now be derived
in exactly the same way as~(\ref{qeq}) was, and
regarded as not just a conjugate clown of the
latter but as an equal dual equation.
\subsection{Baxter equation}
It reads
\[ t_\tau(\lambda)Q(\lambda)=Q(\lambda+{\tfrac12})
+(e^{4\pi i\tau\lambda}+1)^N Q(\lambda-{\tfrac12}) , \]
where~\(t_\tau(\lambda)\) and~\(Q(\lambda)\) are two families of
elements (of~\(\mathcal A_\tau\) and~\(\mathcal B\) respectively),
which all commute with each other
\[ [t_\tau(\lambda),t_\tau(\mu)]=[Q(\lambda),
t_\tau(\mu)]=[Q(\lambda),Q(\mu)] .  \] In this paper we do not define
those families explicitly and do not derive the equation.  Let us only
mention that the chiral evolution operator~\(Q\) is in fact
~\(Q(\lambda)\) evaluated at a particular value of $\lambda$. The
duality symmetry of our construction implies that there is another
dual equation having the form
\[ t_{\frac1\tau}(\lambda)Q(\lambda)=Q(\lambda
+{\tfrac\tau2})+(e^{4\pi i\lambda}+1)^N Q(\lambda-{\tfrac\tau2}). \]
Thus, the Baxter equation gets accociated with the modular lattice.

We stop here and postpone the study of Baxter equations to the next
paper of this series.

\section{Conclusions and comments}

We have shown, that the appropriate lattice regularization of the
quantum Liouville model allows to enter the ``forbidden region'' of
the Virasoro central charge $1\le\centc\le25$. The key for this is
using the double family of dual quantum fields, which are not
selfadjoint, but normal and adjoint to each other. The results are
still rather modest, but proving the point quite persuasively.

The real problem is that of findng the spectrum of the model. One
viable approach to this can be based on the Baxter equation. In the
next paper of this series we plan to deal with this equation. The idea
of its derivation originally goes to Baxter himself \cite{Bax8v}. In
\cite{BazStr,BaxBazPer} it was realized in the context of the chiral
Potts model, and in \cite{BLZ1,BLZ2,BLZ3} in the context of quantum
conformal field theory. The dual Baxter equations in a different
context were also considered recently by Smirnov \cite{Smir}.

In parallel construction for the WZNW model \cite{faddeev91} it was
shown how to separate the zero mode problem which reduces the problem
of finding the spectrum of the primary states to a problem with finite
number of degrees of freedom.  It is possible that similar
construction can be found here. If so, one will be able to bypass
Baxter equations for the investigation of the spectrum.

Liouville model is known to be a contraction of the massive
Sine--(Sinh--) Cordon model to which most of our considerations are
also applicable \cite{fadVol94}.

In a series of papers \cite{BLZ1,BLZ2,BLZ3} the quantum KdV equation
was considered without recourse to lattice. As the Liouville and KdV
models are close relatives, it is no wonder that one can see many
similarities between those papers and our text. Moreover, the discrete
variant of quantum KdV was already considered in \cite{volkov97b}.  We
believe, that using lattice allows to use the duality in full strength
and, in particular, go to the strong coupling region.

In papers \cite{gervais1,gervais3} arguments were raised for the
exceptional values of the central charge $\centc=7,\ 13,\ 19$. From
our point of view, these values are distinguished only by giving our
modular lattice supplementary symmetries ($\tau$ corresponds to
elliptic fixed points of the modular figure: $ e^{\IMUN\pi/3},\ 
e^{\IMUN\pi/2},\ e^{\IMUN 2\pi/3}$).  There is no reason for us to
exclude other values of $\tau$ on the unit half-circle.

Finally, we cannot help stating that for us the function $f(z)$
defined in eqn~(\ref{eq:fofz}) seems to be the real cornerstone of the
theory of quantum integrable models. It appears in basic objects of
the dynamical theory: the evolution and the Baxter operators. Its
close relative as a function of rapidity defines the Zamolodchikov's
factorized S-matrices. It is also indispensable in the theory of
form-factors \cite{smirnov}. We believe that the full content of
duality, hidden in this function, is still not explored to full
extent. In the Appendix below we describe some of its remarkable
properties.

\section{Appendix: the non-compact quantum dilogarithm}

Let complex $\la$ have a nonzero real part $\Re \la\ne0$.  The
non-compact QDL, $\QDILOG(z)$, $z\in\COMPLEXS$, $|\Im z|<|\Im\CLA|$,
\[
\CLA\equiv\IMUN(\la+\la^{-1})/2,
\] 
is defined by the formula
\begin{equation}\label{ncqdl}
\QDILOG(z)\equiv\exp\left(\frac{1}{4}
\int_{-\infty}^{+\infty}
\frac{e^{-2\IMUN z x}\, dx}{\sinh(x\la)
\sinh(x/\la) x}\right),
\end{equation}
where the singularity at $x=0$ is put below the contour of
integration.  This definition implies that $\QDILOG(z)$ is unchanged
under substitutions $\la\to\la^{-1}$, $\la\to-\la$. Using this
symmetry, we choose $\la$ to lay in the first quadrant of the complex
plane, namely
\[
\Re\la>0,\quad \Im\la\ge 0,
\]
which implies that
\[
\Im\tau>0,\quad\tau\equiv\la^2.
\]
This function has the following properties.

\subsection{Functional relations}
Function (\ref{ncqdl}) satisfies the `inversion' relation
\begin{equation}\label{inversion}
\QDILOG(z)\QDILOG(-z)=e^{\IMUN\pi z^2
-\IMUN\pi(1+2\CLA^2)/6},
\end{equation}
and a pair of functional equations
\begin{equation}\label{shift}
\QDILOG(z-\IMUN\la^{\pm1}/2)=(1+e^{2\pi z\la^{\pm1} })
\QDILOG(z+\IMUN\la^{\pm1}/2).
\end{equation}
The latter equations enable us to extend the definition of the QDL to
the entire complex plane.

When $\la$ is real or a pure phase, function $\QDILOG(z)$ is {\em
  unitary} in the sense that
\begin{equation}\label{qdlunitarity}
\overline{\QDILOG(z)}=1/\QDILOG(\bar z).
\end{equation}
If selfadjoint operators $\MOM$ and $\POS$ in $L^2(\REALS)$ satisfy
the Heisenberg commutation relations
\begin{equation}\label{heisen}
[\MOM,\POS]=\frac{1}{2\pi \IMUN},
\end{equation}
the following operator five term identity holds:
\begin{equation}\label{pent}
\QDILOG(\MOM)\QDILOG(\POS)=\QDILOG(\POS)\QDILOG(\MOM+\POS)
\QDILOG(\MOM).
\end{equation}
For real $\la$ this can be proved in the $C^*$-algebraic framework
\footnote{S.L. Woronowicz: private communication, 1998}. We will prove
it for complex $\la$. The case of real $\la$ then will follow by
continuity.
\subsection{Analytic properties }
We can perform the integration in (\ref{ncqdl}) by the residue method.
The result can be written as ratio of two q-exponentials
\begin{equation}\label{ratio}
\QDILOG(z)=
(e^{2\pi (z+\CLA)\la};\QPAR^2)_\infty/
(e^{2\pi (z-\CLA)\la^{-1}};\QTIL^2)_\infty,
\end{equation}
where
\[
\QPAR=e^{\IMUN\pi\la^2},\quad \QTIL=e^{-\IMUN\pi\la^{-2}},
\]
thus reprpoducing definition~(\ref{eq:fofz}) in the text.
Formula~(\ref{ratio}) defines a meromorphic function on the entire
complex plane, satisfying functional equations (\ref{inversion}) and
(\ref{shift}), with essential singularity at infinity.  So, it is the
analytical continuation of definition~(\ref{ncqdl}) to the entire
complex plane.  It is easy to read off location of its poles and
zeroes:
\[
\mathrm{zeroes\ of\ }(\QDILOG(z))^{\pm1}=
\{\mp(\CLA+m\IMUN\la+n\IMUN\la^{-1}):\ m,n\in\INTEGERS_{\ge0}\}.
\]  
The behavior at infinity depends on the direction along which the
limit is taken:
\begin{equation}\label{asymp}
\QDILOG(z)\bigg\vert_{|z|\to\infty} \approx\left\{
\begin{array}{ll}
1&|\arg(z)|>\frac{\pi}{2}+\arg(\la);\\
e^{\IMUN\pi z^2-\IMUN\pi (1+2\CLA^2)/6}&|\arg(z)|<\frac{\pi}{2}-\arg(\la);\\
(\bar q^2;\bar q^2)_\infty/\Theta(\IMUN\la^{-1}z;-\la^{-2})&
|\arg z-\pi/2|<\arg\la;\\
\Theta(\IMUN\la z;\la^{2})/(q^2; q^2)_\infty&
|\arg z+\pi/2|<\arg\la,
\end{array}\right.
\end{equation}
where
\[
\Theta(z;\tau)\equiv\sum_{n\in\INTEGERS}e^{\IMUN\pi\tau n^2+2\pi\IMUN
  nz}, \quad\Im\tau>0.
\]
Thus, for complex $\la$, double quasi-periodic $\theta$-functions,
generators of the field of meromorphic functions on complex tori,
describe the asymptotic behavior of the non-compact QDL.

\subsection{Integral Ramanujan identity}

Consider the following Fourier integral:
\begin{equation}\label{ramanint}
\RAMAN(u,v,w)\equiv
\int_{\REALS}\frac{\QDILOG(x+u)}{\QDILOG(x+v)}e^{2\pi\IMUN wx}\, dx,
\end{equation}
where
\begin{equation}\label{restrictions1}
\Im(v+\CLA)>0,\quad\Im(-u+\CLA)>0, \quad \Im(v-u)<\Im w<0.
\end{equation}
Restrictions (\ref{restrictions1}) actually can be considerably
relaxed by deforming the integration path in the complex $x$ plane,
keeping the asymptotic directions of the two ends within the sectors
$\pm(|\arg x|-\pi/2)>\arg\la$. So, the enlarged in this way domain for
the variables $u,v,w$ has the form:
\begin{equation}\label{restrictions2}
|\arg (\IMUN z)|<\pi-\arg\la,\quad z\in\{w,v-u-w,u-v-2\CLA\}.
\end{equation}
Regarding $\QDILOG(z)$ as a `non-compact' analogue of the q-exponent
$(x;q)_\infty$, definition (\ref{ramanint}) can be interpreted as the
corresponding integral counterpart of the Ramanujan sum:
\[
{}_1\!\psi_1(x,y,z)\equiv\sum_{n\in\INTEGERS}\frac{(x;q)_n}{(y;q)_n}z^n.
\]
The latter is known to be evaluated explicitly, the result being the
famous Ramanujan summation formula:
\[
{}_1\!\psi_1(x,y,z)=\frac{(q;q)_\infty(y/x;q)_\infty(xz;q)_\infty
  (q/xz;q)_\infty}{(y;q)_\infty(q/x;q)_\infty(z;q)_\infty(y/xz;q)_\infty}.
\]
Remarkably, integral (\ref{ramanint}) can be evaluated explicitly as
well. Indeed, using the residue method, we easily come to the
following result:
\begin{gather}\label{ramanres1}
  \RAMAN(u,v,w)=
  \frac{\QDILOG(u-v-\CLA)\QDILOG(w+\CLA)}{\QDILOG(u-v+w-\CLA)}
  e^{-2\pi\IMUN w(v+\CLA)+\IMUN\pi(1-4\CLA^2)/12} \\\label{ramanres2}
  =\frac{\QDILOG(v-u-w+\CLA)}{\QDILOG(v-u+\CLA)\QDILOG(-w-\CLA)}
  e^{-2\pi\IMUN w(u-\CLA)-\IMUN\pi(1-4\CLA^2)/12},
\end{gather}
where the two expressions in the right hand side are related to each
other through the inversion relation (\ref{inversion}). The similarity
of this result with Ramanujan sum becomes very transparent if we
rewrite the latter in the form:
\begin{equation}\label{ramsum}
\sum_{n\in\INTEGERS}\frac{(yq^n;q)_\infty}{(xq^n;q)_\infty}z^n=
\frac{(y/x;q)_\infty(q/z;q)_\infty
}{(y/xz;q)_\infty}\frac{\theta_q(xz)}{\theta_q(x)\theta_q(z)}(q;q)^2_\infty,
\end{equation}
where the $\theta_q$-function is defined by
\begin{equation}\label{raminv}
\theta_q(x)\equiv (q;q)_\infty(x;q)_\infty(q/x;q)_\infty=
\sum_{n\in\INTEGERS}q^{n(n-1)/2}(-x)^n.
\end{equation}
Comparing the inversion relation (\ref{inversion}) with
eqn~(\ref{raminv}), we conclude that the non-compact analogue of the
$\theta$-function is the Gaussian exponent, and the structures of
eqns~(\ref{ramsum}) and (\ref{ramanres1}) are now quite similar.

\subsection{Fourier transformation of the QDL}

Certain specializations of $\RAMAN(u,v,w)$ lead to the following
Fourier transformation formulas for the QDL:
\begin{multline}\label{fourier1}
  \lefteqn{\FQDILOG_+(w)\equiv \int_{\REALS}\QDILOG(x)e^{2\pi \IMUN
      wx}\,dx
    =\RAMAN(0,v,w)\vert_{v\to-\infty}}\\
  =e^{2\pi \IMUN w\CLA-\IMUN\pi(1-4\CLA^2)/12}/
  \QDILOG(-w-\CLA)=e^{-\IMUN\pi w^2+\IMUN\pi(1-4\CLA^2)/12}
  \QDILOG(w+\CLA),
\end{multline}
and
\begin{multline}\label{fourier2}
  \lefteqn{\FQDILOG_-(w)\equiv \int_{\REALS}(\QDILOG(x))^{-1}e^{2\pi
      \IMUN wx}\,dx
    =\RAMAN(u,0,w)\vert_{u\to-\infty}}\\
  =e^{-2\pi\IMUN w\CLA+\IMUN\pi(1-4\CLA^2)/12} \QDILOG(w+\CLA)=
  e^{\IMUN\pi w^2-\IMUN\pi(1-4\CLA^2)/12}/\QDILOG(-w-\CLA),
\end{multline}
The corresponding inverse transformations read:
\begin{equation}\label{finv}
(\QDILOG(x))^{\pm1}=\int_{\REALS}\FQDILOG_\pm(y)e^{-2\pi\IMUN xy}dy,
\end{equation}
where the pole at $y=0$ is surrounded from below.

\subsection{Proof of the Pentagon identity}

Using formula (\ref{finv}) and commutation relation (\ref{heisen}), we
equate the coefficients of the operator terms
\[
e^{-2\pi\IMUN x \POS }e^{-2\pi\IMUN y \MOM}
\]
in the pentagon relation (\ref{pent}), the result being an integral
identity:
\[
\FQDILOG_+(x)\FQDILOG_+(y)e^{2\pi\IMUN x y}=
\int_{\REALS}\FQDILOG_+(z)\FQDILOG_+(x-z)\FQDILOG_+(y-z)e^{\IMUN\pi
  z^2}dz,
\]
where the singularities at $z=x$, $z=y$ are put below, and at $z=0$,
above the integration path.  Now, multiplying both sides of this
identity by $\exp(-2\pi\IMUN y u)$, integrating over $y$, and using
(\ref{finv}), we obtain
\[
\FQDILOG_+(x)\QDILOG(u-x)=\QDILOG(u)
\int_{\REALS}\FQDILOG_+(z)\FQDILOG_+(x-z)e^{\IMUN\pi z^2-2\pi\IMUN
  uz}dz.
\]
Using (\ref{fourier1}), we rewrite it equivalently
\begin{multline}
  \lefteqn{\frac{\QDILOG(u-x)}{\QDILOG(-x-\CLA)\QDILOG(u)}
    e^{-\IMUN\pi(1-4\CLA^2)/12}}\\
  =\int_{\REALS}\frac{\QDILOG(z+\CLA)}{\QDILOG(z-x-\CLA)}
  e^{-2\pi\IMUN z(u+\CLA)}dz=\RAMAN(\CLA,-x-\CLA,-u-\CLA),
\end{multline}
which is a particular case of (\ref{ramanres2}).

\end{document}